\documentclass[fleqn,usenatbib]{mnras}

\usepackage{newtxtext,newtxmath}
\usepackage[T1]{fontenc}

\DeclareRobustCommand{\VAN}[3]{#2}
\let\VANthebibliography\thebibliography
\def\thebibliography{\DeclareRobustCommand{\VAN}[3]{##3}\VANthebibliography}

\usepackage{graphicx}	
\usepackage{amsmath}	
\usepackage{mathtools}

\usepackage{multirow}
\usepackage{array}
\newcolumntype{L}[1]{>{\raggedright\let\newline\\\arraybackslash\hspace{0pt}}m{#1}}
\newcolumntype{C}[1]{>{\centering\let\newline\\\arraybackslash\hspace{0pt}}m{#1}}

\newcommand{\pcm}{\,cm$^{-2}$}	
\newcommand{\erg}{erg cm$^{-2}$ s$^{-1}$}

\newcommand{\astro}{\emph{AstroSat}}
\newcommand{\src}{SGR J1830$-$0645}

\title[AstroSat observation of \src]{\astro\ observation of the magnetar \src\ during its first detected X-ray outburst}

\author[R. Sharma et al.]{
Rahul Sharma,$^{1}$\thanks{E-mail: rahul1607kumar@gmail.com; rsharma@rri.res.in (RS)}
Chetana Jain,$^{2}$\thanks{E-mail: chetanajain11@gmail.com (CJ)}
Biswajit Paul$^{1}$
and T. R. Seshadri$^{3}$
\\
$^{1}$Raman Research Institute, C.V. Raman Avenue, Sadashivanagar, Bengaluru 560080, Karnataka, India \\
$^{2}$Hansraj College, University of Delhi, Delhi 110007, India\\
$^{3}$Department of Physics and Astrophysics, University of Delhi, Delhi 110007, India\\
}

\date{Accepted XXX. Received YYY; in original form ZZZ}

\pubyear{2023}

\begin{document}

\label{firstpage}
\pagerange{\pageref{firstpage}--\pageref{lastpage}}
\maketitle

\begin{abstract}

We present here timing and spectral analyses of \src\ based on an \astro\ observation carried out on 2020 October 16, about a week after the onset of its first detected X-ray outburst. Using data taken with the Soft X-ray Telescope (SXT) and Large Area X-ray Proportional Counter (LAXPC), we have detected 0.9--10 keV coherent pulsations at a period of $\sim$10.4 s. The pulse profiles were single-peaked, asymmetric and consisted of minor peaks attributable to hotspots on the neutron star surface. The pulsed fraction evolved significantly with energy, increasing to energies around 5 keV with a steep drop thereafter. The 0.9--25 keV SXT--LAXPC energy spectrum is best described with two thermal components having temperatures $\sim$0.46 and $\sim$1.1 keV (emission radii of $\sim$2.4 and $\sim$0.65 km, respectively, assuming a distance of 4 kpc) along with a power-law component having a photon index of $\sim$0.39. We report the detection of 67 X-ray bursts having an average duration of $\sim$33 ms. The brightest burst lasted for about 90 ms and had a 3--25 keV fluence of  $\sim 5 \times 10^{-9}$ erg \pcm.

\end{abstract}




\section{Introduction}

Magnetars are isolated neutron stars that are powered by the decay of their ultrastrong surface dipolar magnetic field ($\sim$10$^{15}$ G) \citep{Duncan92, Thompson96}. These ultrahigh magnetic field objects are relatively young and their dynamic magnetosphere is endowed with strong temporal variability which typically includes, a slow rotation period (2--12 s), a rapid spin-down on a time-scale of a few thousand years, bright short millisecond to seconds bursts and month-long outbursts \citep{Collazzi15, Kaspi17}.
The X-ray pulse profiles are generally very broad, with a strong energy dependence of pulsed fraction. During an outburst, the source X-ray flux is known to increase by $\sim$2--3 orders of magnitude \citep[e.g.,][]{Rea11, Coti18}. The outburst decay time to the quiescent level occurs on a time-scale ranging from a few weeks to several years. 

The persistent emission from magnetars is often parametrized by a blackbody component (kT$\approx$0.3-0.5 keV) (often double blackbody) plus a power-law component ($\Gamma \sim$2--4) \citep{Olausen14}. The magnetar spectra are seen to be highly spin-phase dependent, an observation that is corroborated by the strong energy dependence of the pulse profiles \citep{den08a, den08b}. The spectra are known to harden during an outburst and gradually soften during its decay. In some relatively low magnetic field magnetars ($B \sim 10^{14}$ G), such as SGR 0418+5729 and Swift J1822.3--1606, pulse-phase dependent absorption lines have been reported, which are interpreted to be due to proton cyclotron resonant scattering \citep{Tiengo13, Rod16}. 

In this work, we present the timing and spectral analyses of \src\ which displays characteristics typical of the bulk of the magnetar family. \src\ was discovered with the Burst Alert Telescope (BAT) onboard the Neil Gehrels \textit{Swift} Observatory (hereafter \textit{Swift}) on 2020 Octobser 10 following a soft, short gamma-ray burst from its direction \citep{Page20}. Subsequent observations with \textit{Swift} - X-ray Telescope (XRT) revealed the rotation properties of \src\ \citep{Gogus20a}. 
This source exhibited coherent pulsation at a frequency of $\sim$ 0.096 Hz. The spin period was later confirmed by \citet{Ray20} and \citet{Younes20} from observations made with the 0.2--12 keV X-ray telescope onboard the Neutron Star Interior Composition Explorer (\textit{NICER}). Using the phase-coherent technique, \citet{Coti21} determined a spin-down rate of -6.2$\times$10$^{-14}$ Hz s$^{-1}$ by using data from \textit{Swift}, \textit{XMM--Newton} and \textit{NuSTAR}. Similar results were later confirmed by \citet{Younes22} using NICER observations. The derived spin-down parameters imply a dipole magnetic field strength of $\sim$5.5$\times$10$^{14}$ G at the pole, a spin-down age of $\sim$24 kyr and a spin-down luminosity of $\sim$2.4$\times$10$^{32}$ erg s$^{-1}$  \citep{Coti21, Younes22}.

The energy spectrum of \src\ is well described by a double blackbody model corresponding to a small hot region (kT$_{BB}\sim$1.2 keV) and an extended warm region (kT$_{BB}\sim$0.5 keV) plus a power-law component accounting for non-thermal emission, which dominates above 10 keV \citep{Coti21, Younes22}. The two blackbody emission components do not show any significant variability in temperature throughout the rotation cycle of the pulsar.  
The pulse profiles of the two blackbody emission components are aligned in phase, thereby indicating that the two regions are not spatially separated \citep{Coti21, Younes22b}. The spin modulation pattern observed in the soft X-ray emission is due to changes in the blackbody emission area of both components. Several short X-ray bursts have been observed from NICER and \textit{Swift} \citep{Ray20, Coti21, Younes22}. The spectra of these bursts can be described with either a hot blackbody or a power-law emission.  

For the current work, we have used data from the Soft X-ray Telescope (SXT) and Large Area X-ray Proportional Counter (LAXPC) instruments onboard the \astro\ \citep{Agrawal06, Singh14}. Section §\ref{sec:obs} describes the observation details and the data reduction process. The results from the timing, spectral, and burst analyses of \src\ are presented in Section §\ref{sec:result}. Section §\ref{sec:discuss} discusses the implications of our findings.

\section{Observations}
\label{sec:obs}

India's first multiwavelength astronomical mission \astro\ was launched in 2015 September by the Indian Space Research Organization. It comprises of five scientific instruments which can simultaneously observe a source over a wide energy range from optical to hard X-rays -- Scanning Sky Monitor \citep[]{Ramadevi18}, Ultra-Violet Imaging Telescope \citep[]{Tandon17}, SXT \citep{Singh17}, LAXPCs \citep{Yadav16, Agrawal17} and Cadmium Zinc Telluride Imager \citep[]{Rao17}. 

The data from an \astro\ observation of \src\ (Observation ID T03\_255T01\_9000003922) made on 2020 October 16 were analyzed for this work (see Table \ref{obslog} for observation details). During this observation, \src\ was observed for a span of $\sim$ 260 ks. For the current work, we have used data from SXT and LAXPC only.

\begin{table*}
\caption{Log of \astro\ observation (ID: 9000003922) of \src.}
\centering
\resizebox{1.6\columnwidth}{!}{
\begin{tabular}{c c c c c c c c}
\hline \hline
\astro      &   Start time            & Stop time             & Mode  & Obs span  & Clean exposure\\
instrument  &   (yyyy-mm-dd hh:mm:ss) & (yyyy-mm-dd hh:mm:ss) &       & (ks)      & (ks)\\
\hline
SXT         & 2020-10-16 04:17:32   & 2020-10-19 02:13:32   & PC    & 251.8        & 38.8 \\
LAXPC       & 2020-10-16 03:37:03   & 2020-10-19 04:44:46   & EA    & 263.2        & 108.4 \\
\hline
\end{tabular}}
\label{obslog}
\end{table*}

\subsection{\astro\ - SXT}

\astro\ - SXT is a focusing X-ray telescope which is capable of performing X-ray imaging and spectroscopy in the 0.3--8 keV energy range with an energy resolution of $\sim$150 eV. It consists of a charge-coupled device (CCD) camera which is operated in the Photon Counting (PC) mode with a time resolution of 2.37 s and in the Fast Windowed (FW) mode with a time resolution of 0.278 s. The on-axis effective area is $\sim$90 cm$^2$ at 1.5 keV and the FWHM of the point spread function in the focal plane is $\sim 2$ arcmin \citep[see,][for details]{Singh16, Singh17}.

The SXT data of \src\ were taken in the PC mode. Level-1 data were processed by using \textsc{sxtpipeline} version 1.4b which generated the filtered level-2 cleaned event files. \textsc{sxtevtmergertool} was used to merge the cleaned event files from different orbits in the SXT data. We used \textsc{xselect} v2.4m tool to extract the image, light curve and spectra of \src. A circular region of 15 arcmin radius was considered as a source region around the source location. The ancillary response file (ARF) was created with the \textsc{sxtARFModule} tool by using the ARF file provided by the SXT team. The response file (sxt\_pc\_mat\_g0to12.rmf) and the blank-sky background spectrum file (SkyBkg\_comb\_EL3p5\_Cl\_Rd16p0\_v01.pha) provided by the SXT team were used\footnote{\url{https://www.tifr.res.in/~astrosat_sxt/dataanalysis.html}}. 

\subsection{\astro\ - LAXPC}

\astro\ - LAXPC consists of three co-aligned proportional counters (LAXPC10, LAXPC20 and LAXPC30) covering a broad energy range of 3--80 keV with a total effective area of 6000 cm$^{2}$ at 15 keV \citep{Yadav16, Agrawal17}. Every LAXPC detector has the capability to record the arrival time of photons with a time resolution of 10 $\mu$s. Due to issues related to high background and gain instability of LAXPC10 and detector LAXPC30 being switched off, we have only used data from LAXPC20 for the current work. The energy resolution for LAXPC20 at 30 keV is about 20$\%$ \citep{Antia21}. 

The Event Analysis mode (EA) data from LAXPC20 were used for performing the timing and spectral analyses of \src. The EA mode data were processed by using the LAXPC software\footnote{\url{https://www.tifr.res.in/~astrosat\_laxpc/LaxpcSoft.html}} (\textsc{LaxpcSoft}: version 3.4.3). The light curves and spectra for the source and background were extracted from level-1 files by using the tool \textsc{laxpcl1}. The background in the LAXPC is estimated from the blank-sky observations \citep[for details, see][]{Antia17}. To minimize the background, we have performed all analyses using the data from the top layer (L1, L2) of the LAXPC20 detector \citep{Sharma20, Sharma23b}. We have used corresponding response files to obtain channel-to-energy conversion information while performing energy-resolved analyses. 

The \textsc{as1bary}\footnote{\url{http://astrosat-ssc.iucaa.in/?q=data\_and\_analysis}} tool was used to apply barycentric correction to the photon arrival times in the level-2 files of LAXPC and SXT using the JPL DE405 ephemeris and source position RA (J2000) = 18$^{\rm h}$ 30$^{\rm m}$ 41$^{\rm s}$.64 and Dec. (J2000) = $-$06$^{\rm o}$ 45$'$ 16$''$.9 obtained from Chandra observation \citep{Gogus20, Younes22}.

\section{ANALYSIS AND RESULTS}
\label{sec:result}

\begin{figure}
    \centering
     \includegraphics[width=\columnwidth]{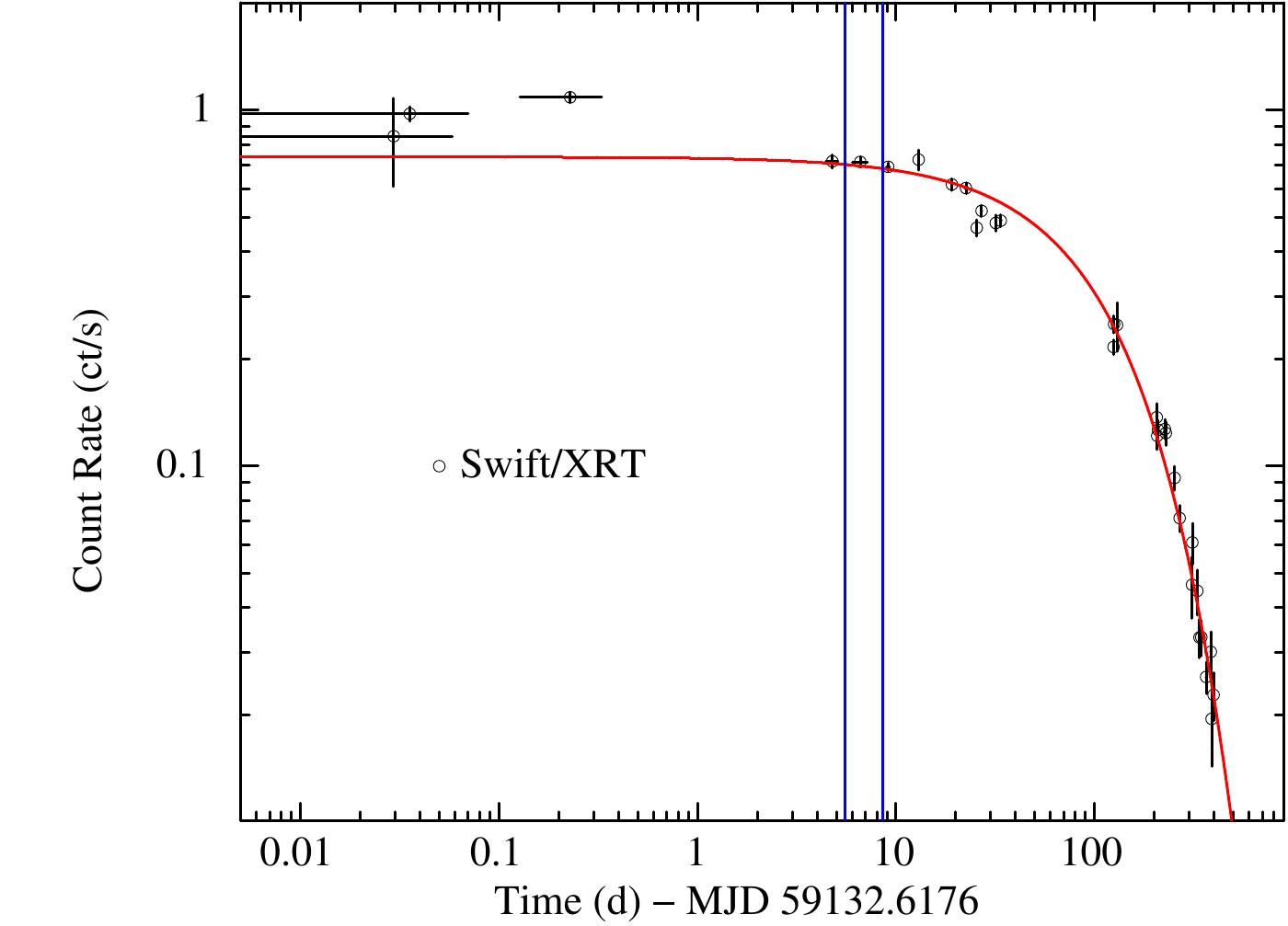}
    \caption{0.5--10 keV \textit{Swift}-XRT light curve of \src\ over a baseline of $\sim$ 400 d since the onset of the outburst (MJD 59132.6176). The red solid line denotes the best-fitting exponential function with an \textit{e}-folding time of 113 $\pm$ 4 d. The vertical blue solid lines mark the observation span of the \astro\ observation. }
    \label{fig:xrt_lc}
\end{figure}

Figure \ref{fig:xrt_lc} shows the evolution of the 0.5–10 keV count rate of \src\ extracted from the 32 observations made with \textit{Swift}-XRT\footnote{Created from the online Build XRT products tool at the UK Swift Science Data Centre.} \citep{Evans07} over a baseline of about 400 d starting from the onset of the outburst (MJD 59132.6176). The time evolution of the count rate can be described with an exponential function ($\propto e^{-t/\tau}$ -- shown by the solid red line) with an \textit{e}-folding time ($\tau$) of 113 $\pm$ 4 days ($\chi^2$ = 221 for 29 degrees of freedom, dof) which is nearly double that reported by \citet{Coti21} using the first 34 d of data coverage of \src.

\subsection{Timing Analysis}

\begin{figure}
    \centering
     \includegraphics[width=\columnwidth]{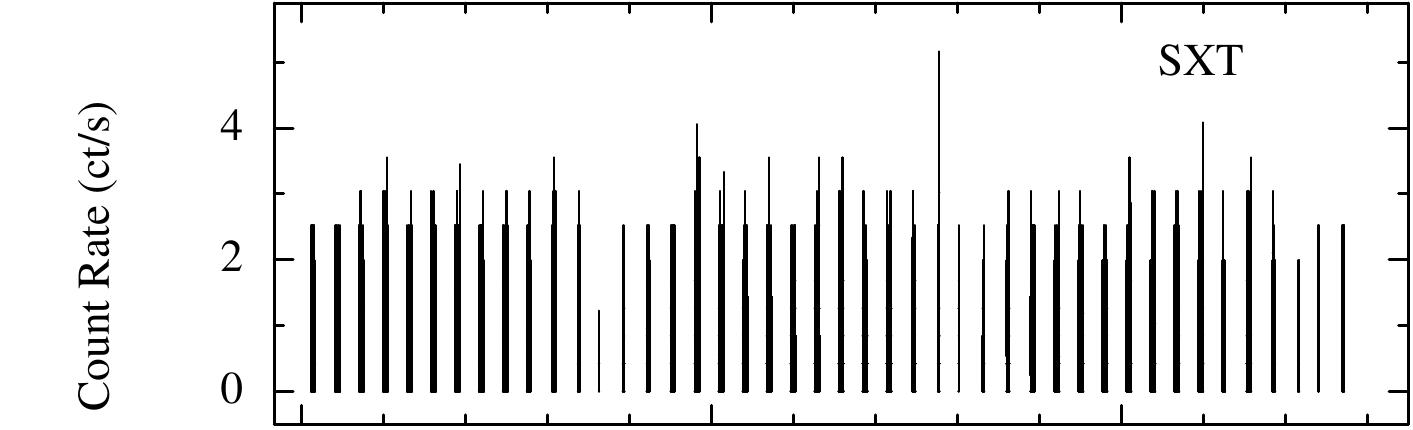}
     \includegraphics[width=\columnwidth]{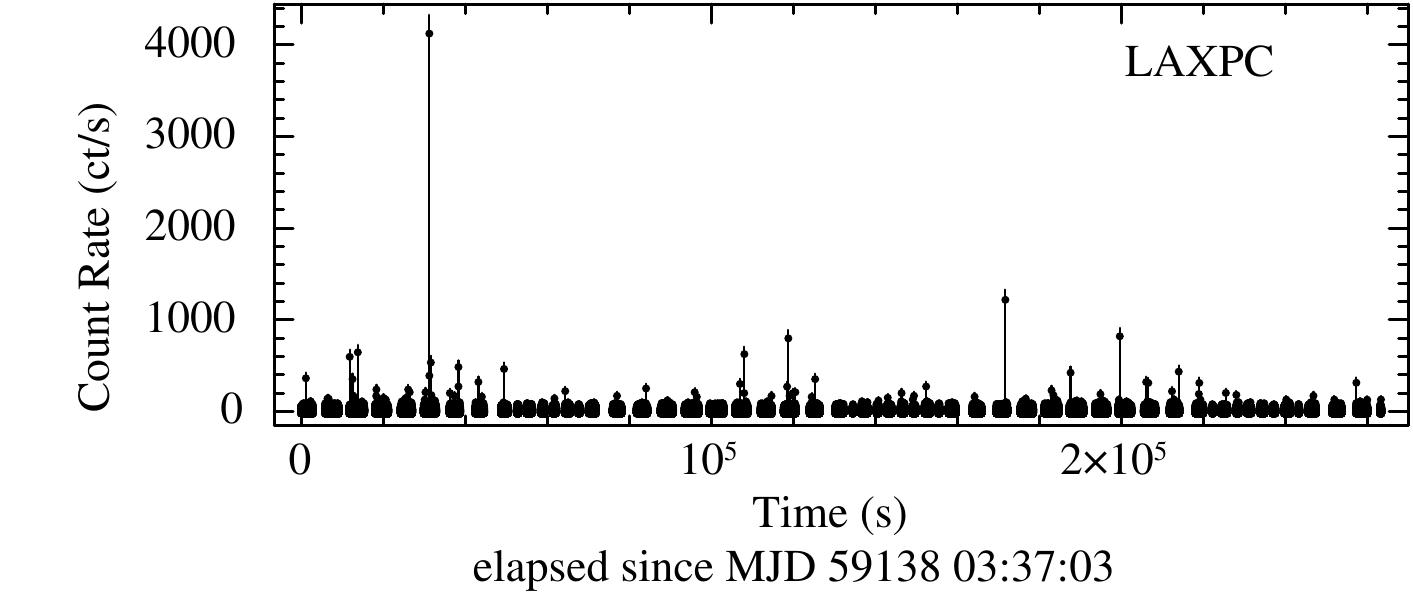}
    \caption{\textit{Top:} The 0.9--7 keV \astro-SXT light curve of \src\ binned at 2.3775 s. \textit{Bottom:} The 3--25 keV \astro-LAXPC light curve of \src\ binned at 0.1 s.}
    \label{fig:laxpc_lc}
\end{figure}

Figure \ref{fig:laxpc_lc} shows the SXT (0.9--7 keV) and LAXPC (3--25 keV) light curves of \src\ binned with 2.3775 s and 0.1 s, respectively. Several short bursts lasting a few milliseconds were detected only with LAXPC during this observation. 

We used the $\chi^2$ maximization technique to determine the spin period of \src. From the literature, the spin period is known to be $\sim$ 10.4157 s. We folded the LAXPC light curve over a range of periods (10.410 -- 10.420 s) with a resolution of $10^{-6}$ s by using the \textsc{efsearch} tool of the XRONOS subpackage of FTOOLS \citep{Blackburn99}. We obtained a spin period of 10.415730 s for epoch MJD 59138. We have detected pulsations only up to 10 keV. The error in the spin period was estimated by using the bootstrap method \citep{Lutovinov12, Boldin13}. We simulated 1000 light curves by the method described in \citet{Sharma22b, Sharma23} and obtained the spin period for each of them by using the epoch-folding technique. We obtained a standard deviation of $4 \times 10^{-6}$ s in the best spin period distribution. This number was taken as the error of the pulse period. From SXT data, we found a spin period of 10.41572 (1) s, consistent with the estimate from the LAXPC data. Using a phase-coherent timing analysis, we have estimated the limit on the spin period derivative to be $\lvert \dot{P} \rvert < 2 \times 10^{-10}$ s s$^{-1}$ at a 90\%  confidence limit.  

\begin{figure}
  \centering
  \includegraphics[width=\linewidth]{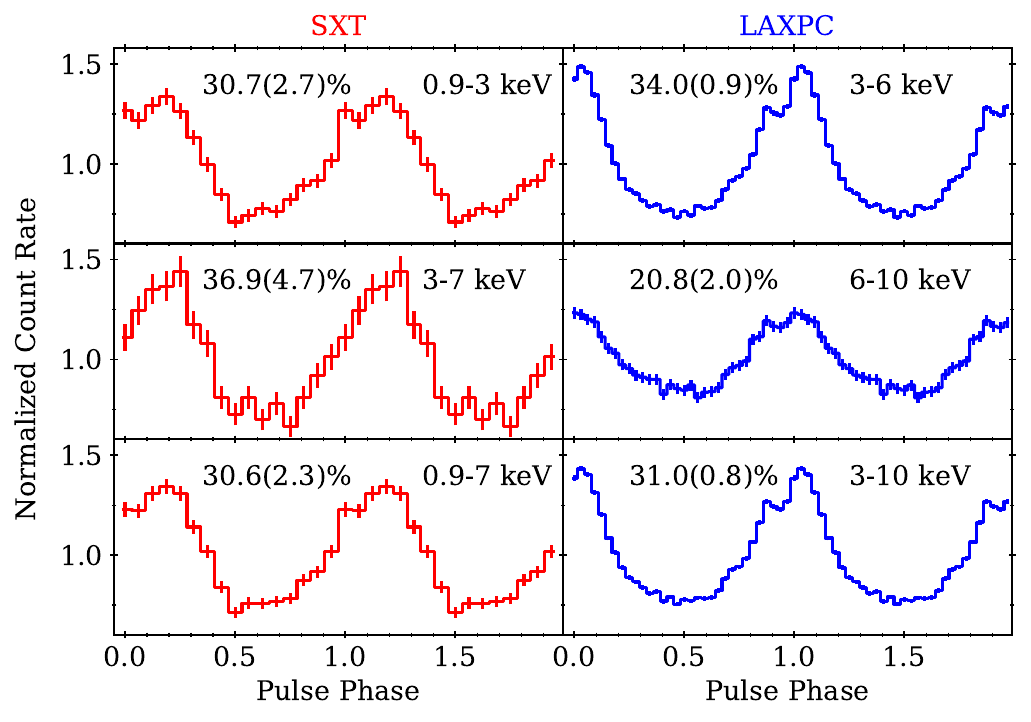} 
    \caption{Energy-resolved pulse profile of \src, generated from SXT (left panels) and LAXPC (right panels). The respective pulsed fraction values are noted in each panel. Two rotation cycles of the pulsar are shown for clarity.}
  \label{fig:pulseprofile}
\end{figure}

\begin{figure}
  \centering
  \includegraphics[width=0.9\linewidth]{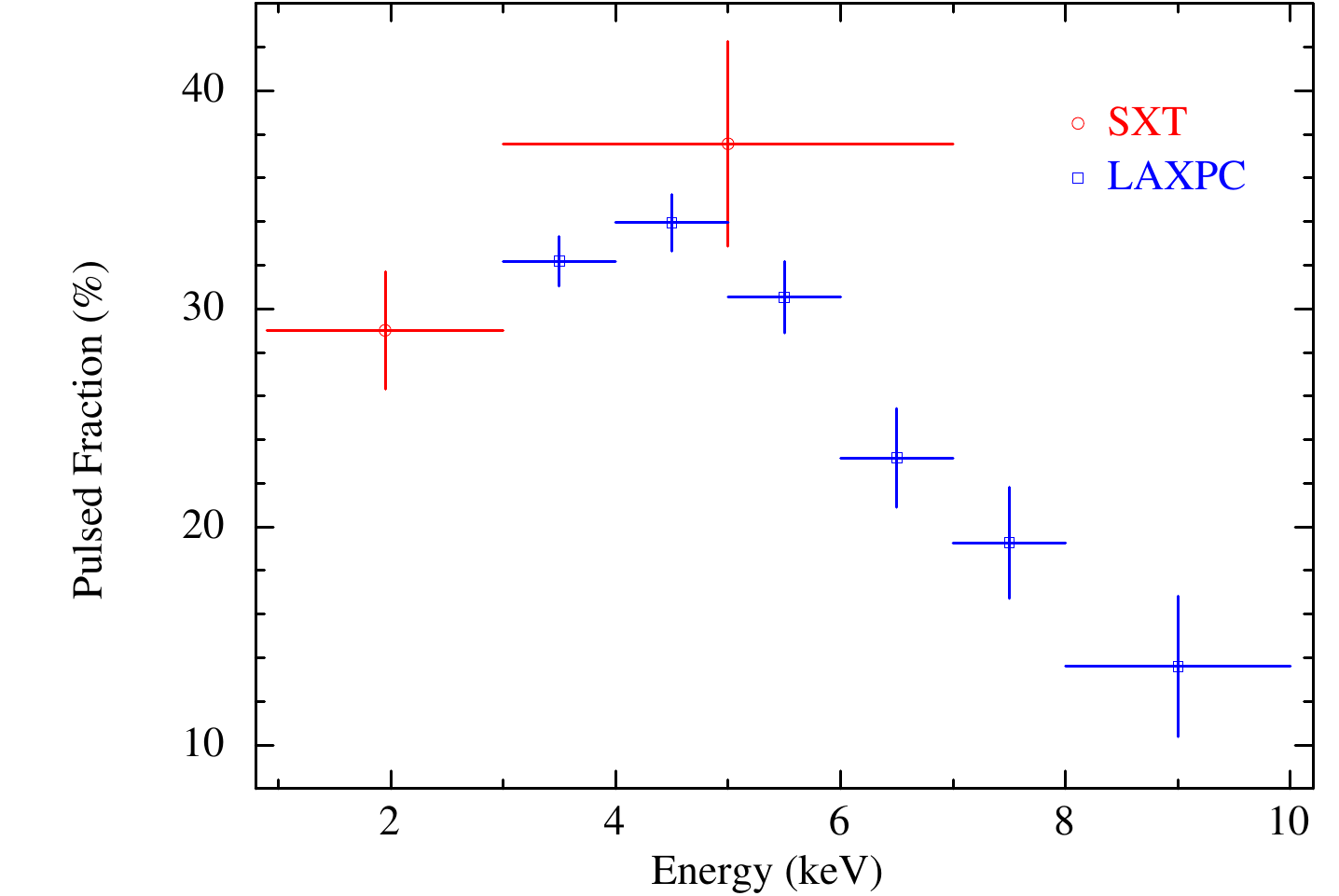}
  \caption{ Evolution of the pulsed fraction in \src\ with energy as seen in the SXT and LAXPC data.}
  \label{fig:pf}
\end{figure}

Figure ~\ref{fig:pulseprofile} shows the energy-resolved pulse profiles of \src, generated from SXT (in the energy ranges 0.9--3, 3--7, and 0.9--7 keV) and LAXPC (in the energy ranges 3--6, 6--10, and 3--10 keV), using the spin period derived with LAXPC. The SXT profiles show a phase shift of $\sim1.4$ s relative to LAXPC, also observed in the pulse profiles reported by \citet{Beri21}. This phase shift could be instrumental and related to the readout time of the SXT CCD.  
The 0.9--3 keV SXT pulse profile displayed an asymmetric single-peak morphology with a minor peak just before the main peak. The 3--7 keV SXT profile was also asymmetric but relatively smoother. The pulsed fraction (defined as the semi-amplitude of the modulation in the pulse profile divided by the average source count rate)\footnote{Pulsed fraction $=\frac{I_{\rm max}-I_{\rm min}}{I_{\rm max}+I_{\rm min}}$, where $I_{\rm max}$ and $I_{\rm min}$ are the maximum and minimum intensities of the pulse profile.} was found to increase from about 30\% at 0.9--3 keV energy to 37\% at 3--7 keV.  
The LAXPC pulse profile displayed a similar morphology, with a minor peak in the rising part of the profile at low energies (3--6 keV), which could possibly be attributed to hotspots on the neutron star surface while the profile in the 6--10 keV is more complex. Similar features were observed in the pulse profiles extracted from the \emph{XMM--Newton} and \emph{NICER} observations \citep{Coti21, Younes22}. 
The LAXPC pulsed fraction was found to decrease from $\sim$34$\%$ (below 6 keV) to $\sim$20$\%$ (at 6--10 keV). Figure ~\ref{fig:pf} shows the significant evolution of the pulsed fraction with energy in \src. It increases up to energies around 5 keV and shows a steep drop thereafter, a behaviour also reported by \citet{Coti21} and \citet{Younes22b}.

\subsection{Spectral Analysis}

We have performed a spectral analysis of the persistent-only emission of \src. We simultaneously fitted spectra from SXT and LAXPC in the 0.9–25 keV energy range to study the spectral properties of the source. We have ignored LAXPC20 data above 25 keV because of the large uncertainty in the background estimation \citep{Antia17, Sharma23}. We have used \textsc{xspec} v12.12.0 from the \textsc{heasoft} 6.29 package for spectral fitting \citep{Arnaud96}. We used a systematic uncertainty of 1\% during spectral fitting \citep{Antia17}. The SXT and LAXPC spectra were grouped using \textsc{grppha} to have a minimum count of 50 counts per bin. A constant factor of 1 fixed for LAXPC was added for cross-calibration.

\begin{figure}
  \centering
  \includegraphics[width=\linewidth]{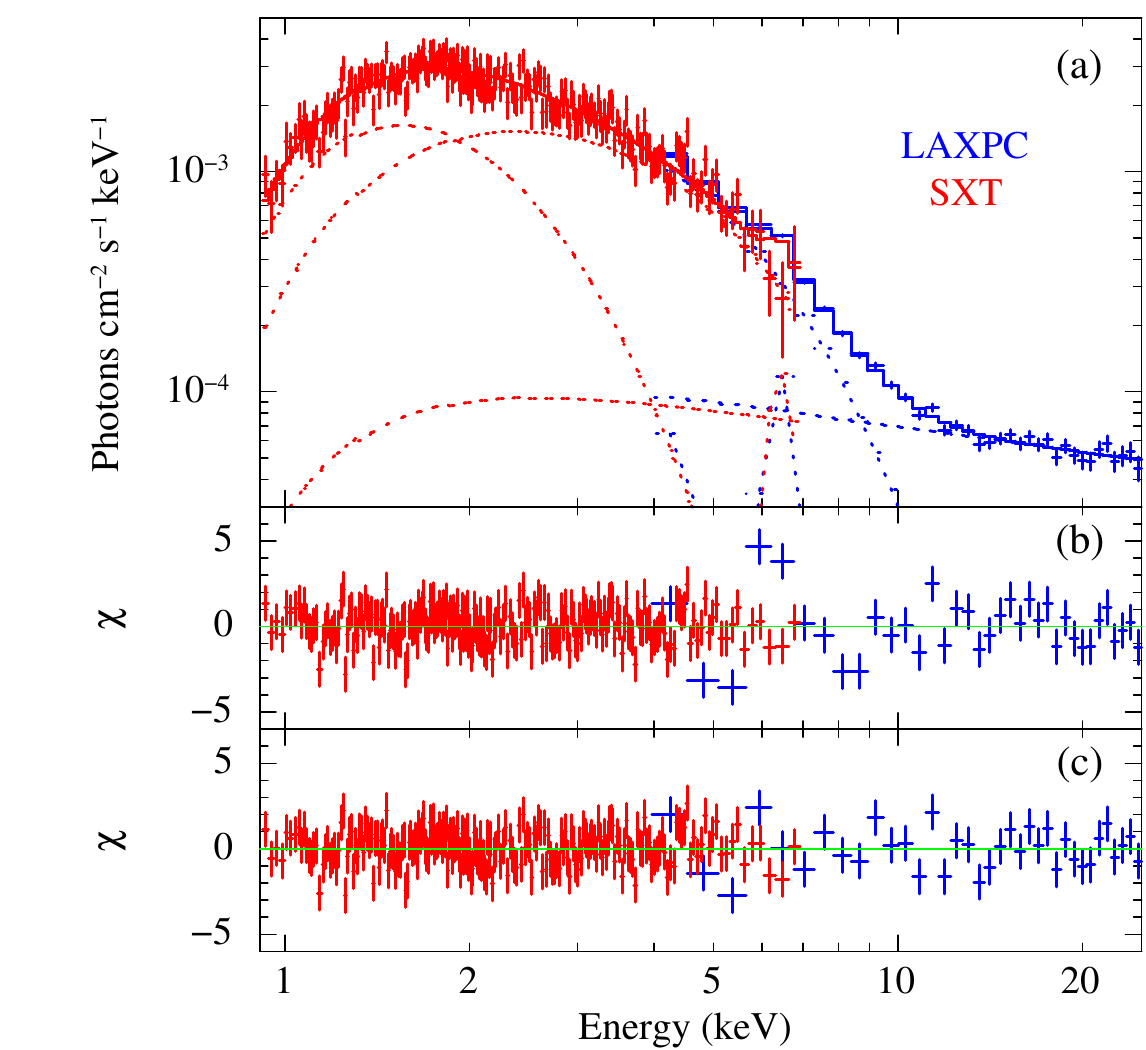}
  \caption{ \textit{Panel (a)}: The 0.9--25 keV energy spectrum of \src\ obtained from SXT and LAXPC data. \textit{(b)}: The residuals for a model comprising continuum emission components without a Gaussian emission line. \textit{(c)}: The residuals with a best-fitting model comprising two blackbodies, a power-law and a Gaussian components.}
  \label{fig:sourcespectrum}
\end{figure}

Figure~\ref{fig:sourcespectrum} shows the 0.9--25 keV energy spectrum of \src\ which can be described with a model comprising a double blackbody along with a power-law component (\texttt{bbodyrad+bbodyrad+powerlaw}) \citep{Coti21}. This model has a $\chi^2$ value of 323 for 253 dof. The addition of a Gaussian feature around 6.4 keV improved the fit ($\chi^2$/dof = 276/250) with an \textit{F}-test probability of improvement by chance of about $10^{-8}$, which corresponds to $5.6 \sigma$ significance. The parameters of the best-fitting spectral model are given in Table~\ref{specs1}. They are consistent with those of \citet{Coti21}, except for the emission line.

\renewcommand{\arraystretch}{1.3}
\begin{table}
\caption{Best-fitting spectral parameters of \src. All errors reported in this table are at a 90\% confidence level ($\Delta \chi^2 = 2.7$).}
\centering
\resizebox{0.8\columnwidth}{!}{
\begin{tabular}{l l r}
\hline \hline
Component & Parameter & Value \\
\hline
Tbabs & $N_H$ ($10^{22}$ cm$^{-2}$) & $1.00^{+0.22}_{-0.19}$ \\[.5ex] 

Bbodyrad  & $kT_{\rm Hot}$ (keV) & $1.096_{-0.04}^{+0.05}$ \\ 
        & $R_{\rm BB}$ (km)$^b$ & $0.65_{-0.07}^{+0.06}$ \\ 
        & norm & $2.66_{-0.56}^{+0.51}$ \\[1ex]

Bbodyrad  & $kT_{\rm Warm}$ (keV) & $0.46_{-0.07}^{+0.09}$ \\ 
        & $R_{\rm BB}$ (km)$^b$ & $2.37_{-0.55}^{+1.25}$ \\
        & norm & $35.1_{-16.4}^{+37.0}$ \\  [1ex]

Powerlaw & $\Gamma$ & $0.39 \pm 0.13$ \\  
    & norm ($10^{-4}$) & $1.73_{-0.5}^{+0.8}$ \\ [1ex]

Gaussian & $E_{\rm line}$ (keV) & $6.44 \pm 0.15$ \\ 
& Sigma (keV) & $0.26_{-0.24}^{+0.25}$ \\  
& norm ($10^{-5}$) & $9.7_{-2.6}^{+3.4}$ \\ 
& EQW (keV) & $0.24^{+0.10}_{-0.09}$ \\[1ex]

Factor & $C_{\rm SXT}$ & $0.91 \pm 0.04$ \\ [1ex] 

        & $F^{\rm BB-hot}_{0.9-25 \rm ~keV}$ & $ 4.06 \times 10^{-11}$ \\ 
        & $F^{\rm BB-warm}_{0.9-25 \rm ~keV}$ & $ 1.40 \times 10^{-11}$ \\ 
Unabs. flux$^a$ & $F^{\rm Powerlaw}_{0.9-25 \rm ~keV}$ & $ 3.03 \times 10^{-11}$ \\ 
        & $F^{\rm Total}_{0.9-25 \rm ~keV}$ & $8.60 \times 10^{-11}$ \\ 
        & $F^{\rm Total}_{0.1-100 \rm ~keV}$ & $3.40 \times 10^{-10}$ \\ 

\hline
    & $\chi^2/{\rm dof}$ & 276.6/250 \\
\hline
\multicolumn{3}{l}{$Note:$ $^a$Flux is in units of \erg.} \\
\multicolumn{3}{L{0.68\columnwidth}}{$^b$Assuming a fiducial distance of 4 kpc \citep{Younes22}.}\\
\end{tabular}}
\label{specs1}
\end{table}

\begin{figure}
  \centering
  \includegraphics[width=\linewidth]{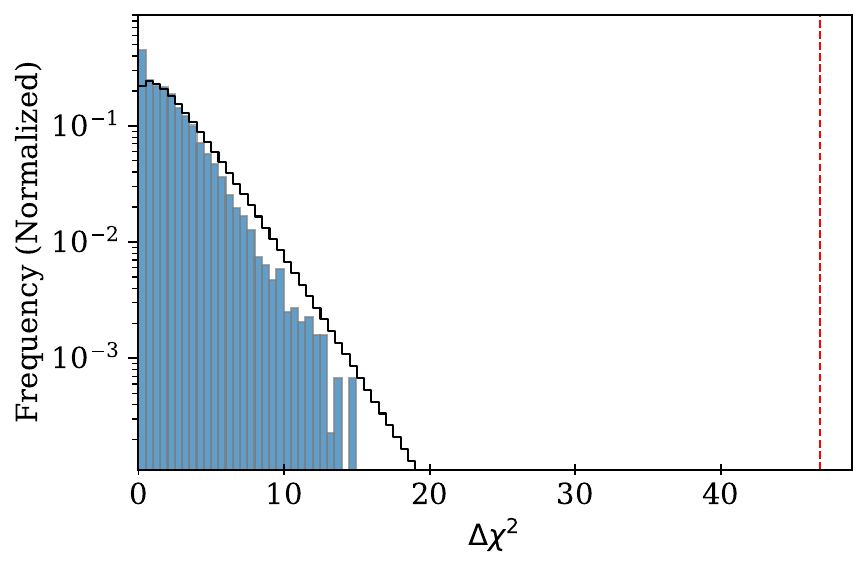}
  \caption{Histogram showing the results of the \texttt{simftest} simulations for testing the detection significance of the emission line. The overlaid solid black curve represents the $\chi^2$ distribution for 3 dof. The vertical red dashed line marks the $\Delta \chi^2$ obtained from spectral fitting of the real data.}
  \label{fig:sim}
\end{figure}

The inclusion of an emission line component in the energy spectrum improved the best-fitting statistics by $\Delta \chi^2$ of 46.8 for 3 additional dof. Although popular amongst the astrophysics community, the \textit{F}-test is associated with caveats when it comes to the detection of emission lines in a spectral model \citep{Protassov02}. Therefore, we estimated the significance of the potential spectral line by using the \texttt{simftest} script from \textsc{xspec}. This routine uses the Monte Carlo method to simulate data sets with the same counting statistics as the original data. We simulated 10,000 data sets, which were fitted with the best-fitting spectral models (both with and without a 6.4 keV emission line). Figure \ref{fig:sim} shows the distribution of $\Delta \chi^2$ values obtained from fitting every simulated data set. From the \texttt{simftest} routine, we obtained a maximum $\Delta \chi^2$ of 15, which is significantly lower than the $\Delta \chi^2$ of 46.8 obtained in the original data. A large deviation of $\Delta \chi^2$ confirms a significant detection of the emission line in the spectra. To be doubly sure, we also tested whether the presence of spectral line is due to systematics in the LAXPC instrument. For this, we increased the systematic error in the spectral fitting to 2\%. We found an improvement of $\Delta \chi^2$ of 20 for 3 additional dof; the \textit{F}-test probability of finding such a change by chance is $\sim 3 \times 10^{-4}$, which corresponds to $3.6 \sigma$ significance. It is important to mention here that the spectral analysis of observations made with the X-ray missions \textit{XMM--Newton} and \textit{NuSTAR} has not detected the presence of this spectral line \citep{Coti21}. Even a dedicated NICER monitoring of the source has not shown any evidence of the presence of  emission lines \citep{Younes22}. Therefore, even though we have performed thorough statistical checks that support the presence of the emission line, it is possible that the 6.4 keV feature observed in our data is an instrumental effect. If our detection of line emission in the persistent emission of \src\ is indeed true, then it has several implications for the X-ray emission mechanism of magnetars, some of which are highlighted in Section~\ref{sec:discuss}.

In order to check the correlation between the emission line and burst/non-burst phase of \src, we generated stacked spectra from 100, 200, 300, 400, and 500 s data before and after each burst. The spectrum from all the segments showed the presence of a $\sim$6.4 keV emission line with a probability of chance improvement $< 10^{-4}$ (i.e. significance $> 3 \sigma$) using \texttt{simftest}, except for the first 100 s segment where we obtained a probability of 0.0029. A relatively high probability, in this case, could be due to overall low exposure ($\sim$ 6.2 ks compared to more than 20 ks for other segments) and thus poorer statistics. The equivalent width of the emission line during these segments was similar to that obtained in the average spectrum of the source ($\sim$0.25--0.30 keV), clearly indicating that the presence of line is not related to the occurrence of bursts in \src.

\subsection{Burst Analysis}

In order to quantify the occurrence of X-ray bursts, we used only the LAXPC data because of the relatively large time resolution (2.37 s) of SXT. We binned the 3--25 keV LAXPC time series with a time resolution of 0.01 s. We searched for bursts in every Good Time Interval (GTI) of the time series by using the \texttt{find\_peaks} routine of the \textsc{scipy} package \citep{Virtanen20}. 

\begin{figure}
  \centering
  \includegraphics[width=\linewidth]{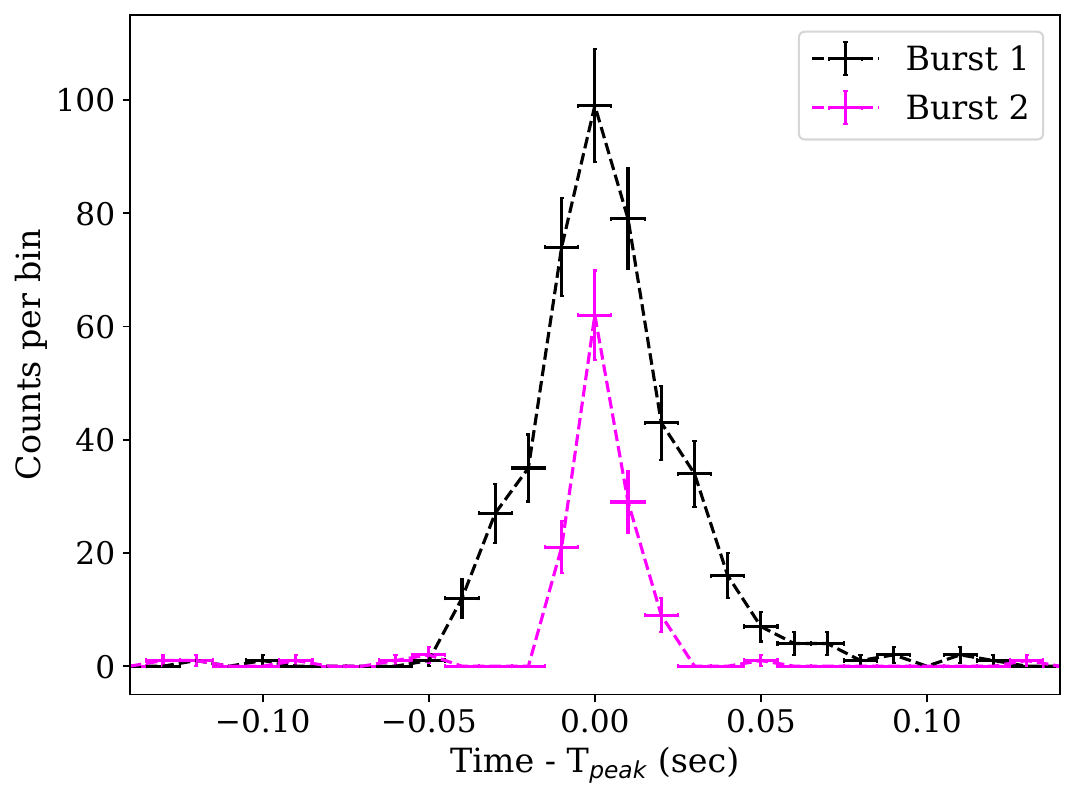}
  \caption{The 3--25 keV profile of the two brightest bursts observed with LAXPC.}
  \label{fig:burstprofile}
\end{figure}

Using a Poisson distribution, the probability ($P_i$) of the number of counts occurring randomly in each time bin ($n_i$) is given by, 
\begin{equation}
    P_i = \frac{\lambda^{n_i} e^{-\lambda}}{n_i !}
\end{equation} 
Here, $\lambda$ is the local mean count rate of every GTI. The events for which $P_i$ was less than 10$^{-4}/N$, were labelled as bursts. Here, $N$ is the total number of time bins in the respective GTI \citep{Gavriil04, Borghese20}. Using this algorithm, we have detected a total of 67 bursts.  Figure \ref{fig:burstprofile} shows the profile of the two brightest bursts observed with LAXPC. The brightest burst had a peak count of about 100 counts in a 10 ms bin, corresponding to a fluence of about 419 counts within a duration of 0.09 s.  

\begin{figure}
  \centering
  \includegraphics[width=\linewidth]{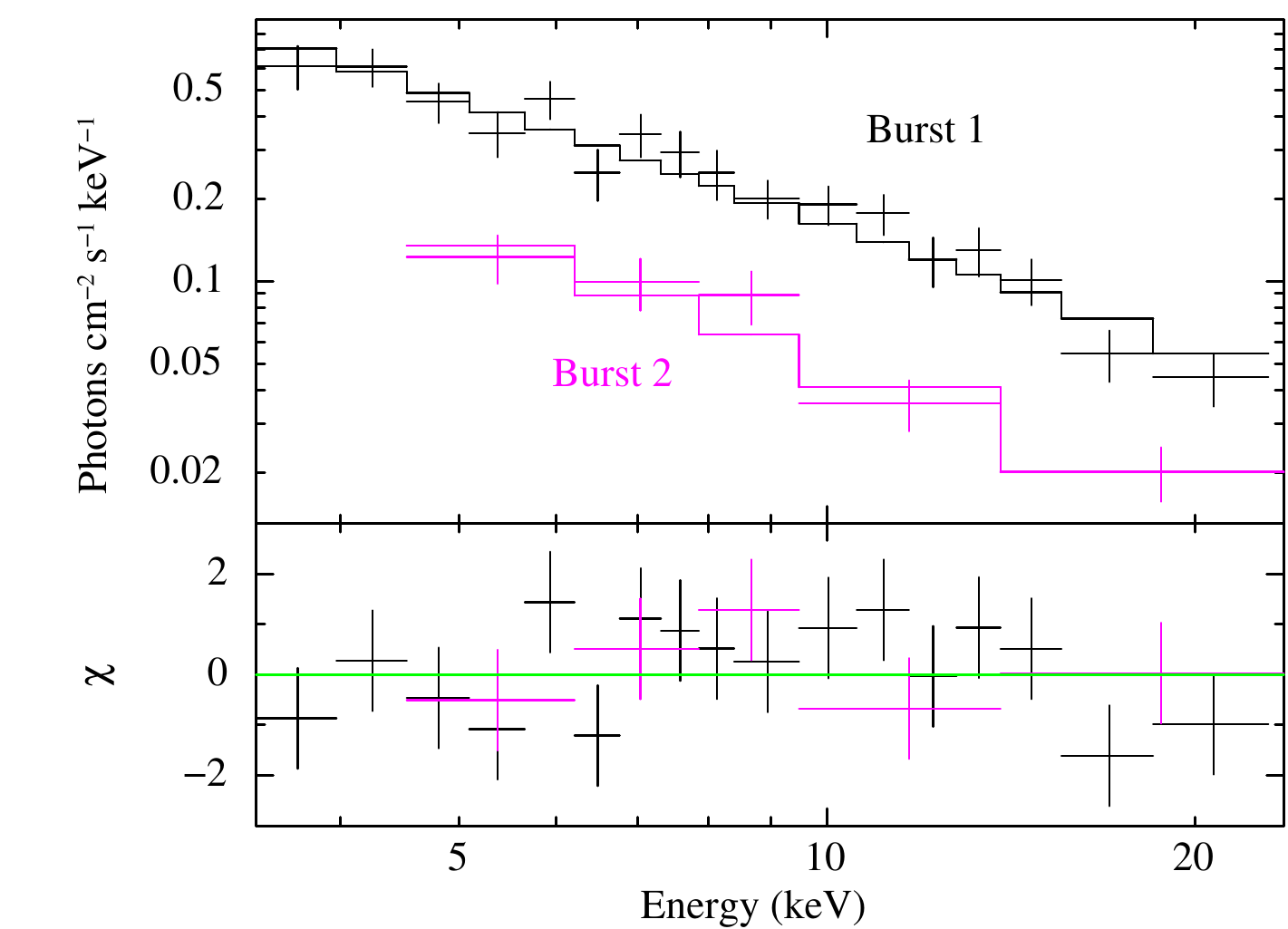}
  \caption{The spectra of the two brightest bursts detected during the \astro\ observation of \src. The bottom panel shows the residuals with respect to the best-fitting absorbed power-law model.}
  \label{fig:burstspectrum}
\end{figure}

Figure \ref{fig:burstspectrum} shows the 3--25 keV burst spectra that have been modelled with an absorbed power-law ($\chi^2/$dof=17.9/18) with $\Gamma=1.52 \pm 0.14$ where $N_H$ was fixed at $1.0 \times 10^{22}$ cm$^{-2}$ (Table \ref{specs2}). For both the bursts, the LAXPC energy spectra could not be modelled with a single blackbody as reported in the literature \citep{Coti21, Younes22}. For the brightest burst, we obtained an unabsorbed 3--25 keV flux of $5.41 \times 10^{-8}$ \erg, corresponding to a fluence of $\sim 5 \times 10^{-9}$ erg \pcm. To date, the \textit{Swift}-BAT burst detected on 2020 November 5 has the highest reported fluence ($\sim3.2\times10^{-8}$ erg \pcm) in 15--150 keV \citep{Coti21}. On extrapolating our results, we obtained a 15--150 keV fluence of $\sim 1.1 \times 10^{-8}$ erg \pcm. The cumulative spectrum of all the 67 bursts is also well described with a power-law ($\Gamma \sim 1.7$) model having an average 3--25 keV unabsorbed flux of $\sim$1$\times$10$^{-10}$ \erg. The burst spectra did not show any evidence of the presence of the $\sim$6.4 keV emission line.

\begin{figure*}
  \centering
  \includegraphics[width=0.47\linewidth]{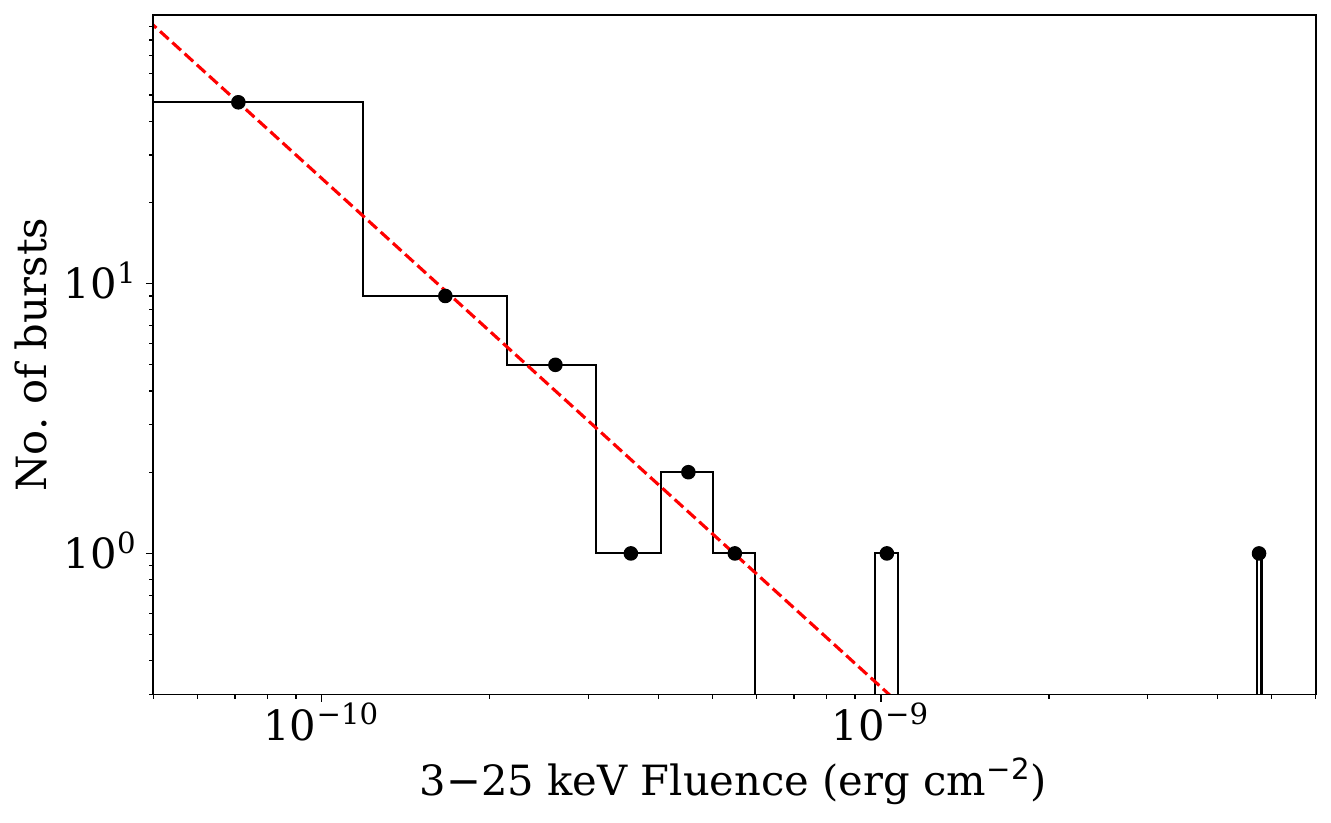}
  \includegraphics[width=0.47\linewidth]{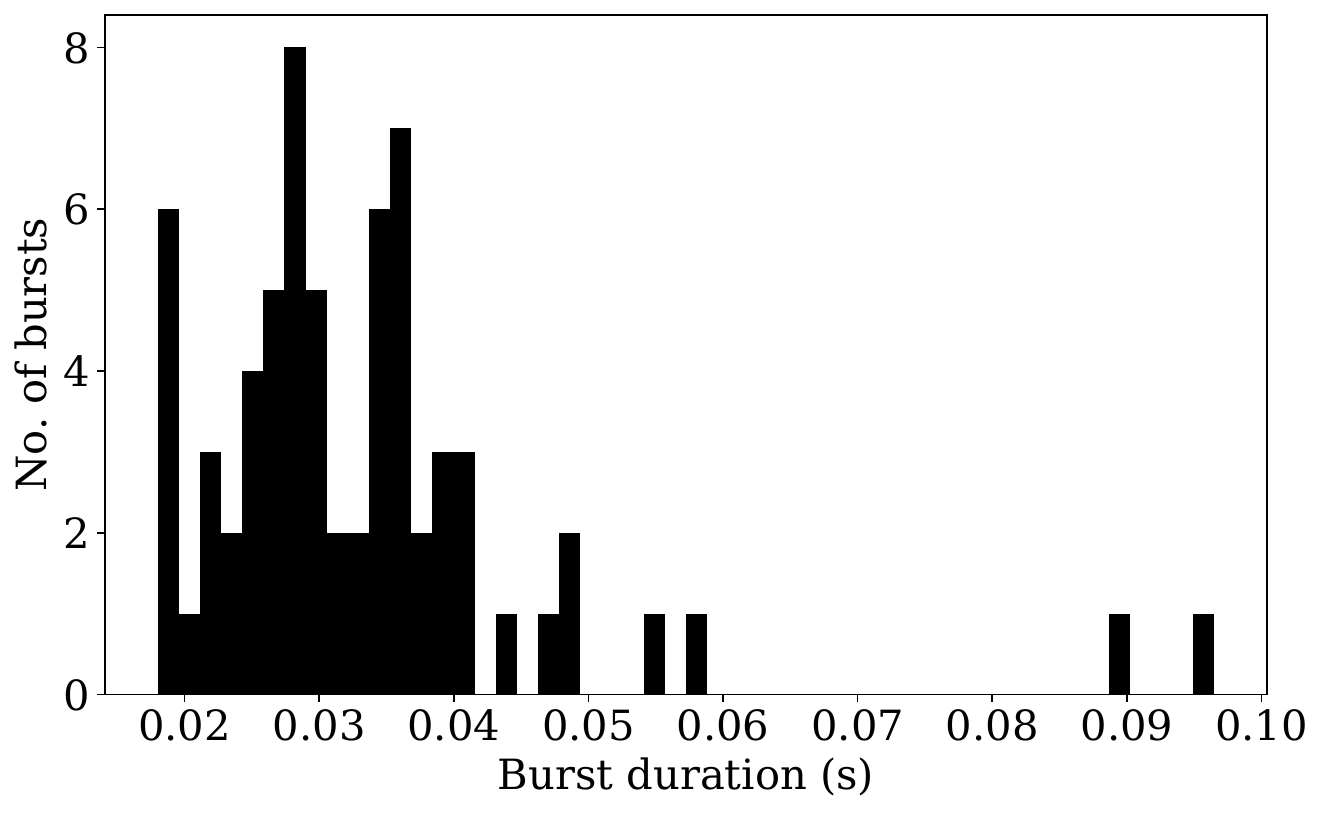}
  \caption{\textit{Left:} The fluence distribution of 3--25 keV bursts detected with \astro\ in \src. The red dashed line shows the best-fitting power-law trend ($F^{-1.88}$). \textit{Right:} Distribution of the duration ($T_{90}$) of the X-ray bursts.} 
  \label{fig:dist}
\end{figure*}

The left panel of Figure~\ref{fig:dist} shows the ﬂuence distribution of all 67 bursts. The 3–25 keV high-fluence tail in this distribution ranges from $(0.2-48) \times 10^{-10}$ erg cm$^{-2}$. This tail can be described by a power-law function having an index of $\sim$1.88. From the 0.7-8 keV NICER data, \citet{Younes22} described the high-fluence tail with a power-law function having an index of 1.5. The right panel of Figure~\ref{fig:dist} shows the distribution of the burst duration ($T_{90}$\footnote{Time interval between 90\% of the peak counts.}). The average duration of all 67 bursts was about 33 ms with a standard deviation of 13 ms.

\renewcommand{\arraystretch}{1.3}
\begin{table}
\caption{Spectral parameters of the two brightest bursts detected in \src\ with LAXPC. The best-fitting model used was the absorbed power-law.}
\centering
\resizebox{\columnwidth}{!}{
\begin{tabular}{l l c c}
\hline \hline
Component & Parameter & Burst 1 & Burst 2 \\[.5ex] 
\hline
Tbabs & $N_H$ ($10^{22}$ cm$^{-2}$) & \multicolumn{2}{c}{$1.0^{\rm fixed}$} \\

Powerlaw & $\Gamma$ & $1.52 \pm 0.14$ & $1.56 \pm 0.38$ \\  
& norm & $5.4_{-1.4}^{+1.9}$ & $1.9_{-1.1}^{+2.4}$\\ 

Unabs. Flux & $F_{3-25 \rm keV}$ (\erg) & $5.41 \times 10^{-8}$ & $1.69 \times 10^{-8}$ \\ 
Unabs. Flux & $F_{0.1-100 \rm keV}$  (\erg) & $1.58 \times 10^{-7}$ & $4.87 \times 10^{-8}$ \\ 

\hline
& $\chi^2/{\rm dof}$ & \multicolumn{2}{c}{17.9/18} \\
\hline
\end{tabular}}
\label{specs2}
\end{table}

\section{DISCUSSION}
\label{sec:discuss}

This work reports results from the timing and spectral analyses of the \astro\ observation of \src\ made on 2020 October 16 during its first detected outburst. We conclude the following from our findings.
\begin{enumerate}
  \item Detection of 0.9--10 keV pulsations in \src. 
  \item Pulse period of 10.415730 (4) s at an epoch of MJD 59138.
  \item Variation in the morphology of the pulse profiles with energy (pronounced dip just before the main peak at low energies and almost sinusoidal at higher energies) along with significant variation in the pulsed fraction.
  \item The 0.9--25 keV SXT+LAXPC energy spectrum comprising the sum of two thermal components and a power-law component.
  \item Tentative detection of a 6.4 keV emission line with an equivalent width of about 0.24 keV.
  \item Detection of several short subsecond X-ray bursts during this observation of \astro.
 \end{enumerate}

The pulsed fraction of \src\ shows a significant evolution with increasing energy. It is observed to increase for energies up to $\sim$5 keV and shows a steep drop thereafter. This trend is different from that observed in several other magnetars, such as 1E 1841-045, 1E 2259+586, and 4U 0142+61 \citep{Kuiper06, An13, Vogel14}. Clearly, the pulsed fraction trend seen in \src\ is in stark contrast to the coronal outflow model of \citet{Beloborodov13}.

\src\ displays spectral properties typical of most magnetars in the soft X-ray band \citep{Thompson02, Tiengo08, Coti18}. The energy spectrum consists of two blackbody (thermal) components along with non-thermal power-law associated with resonantly up-scattered soft thermal photons as they traverse from the stellar surface through the magnetosphere \citep{Fernandez07, Nobili08, Coti21}. Assuming a distance of 4 kpc \citep{Younes22}, we have  estimated the size of the emitting regions to be $R_{\rm BB}$ of 0.65 km (for $kT_{\rm BB}$ of 1.1 keV) and $R_{\rm BB}$ of 2.45 km ($kT_{\rm BB}$ of 0.46 keV). During this observation, the blackbody components carried about 47\% (hot component) and 16\% (warmer component) fractions of the total flux.

We detected a total of 67 bursts from the LAXPC data set with an average duration of 33 ms. The brightest burst lasted for $\sim$90 ms and had a fluence of $\sim 5 \times 10^{-9}$ erg \pcm\ in  the 3--25 keV energy range. The fluence tail of the bursts can be described by a power-law function of index $\sim$1.88. The power-law fluence distribution of burst fluence observed in \src\ is similar to several magnetars  \citep[see e.g.,][]{Cheng96, Scholz11, Collazzi15}  and is believed to be consistent with either magnetospheric reconnection or the crust-quake theories of the burst-triggering mechanism \citep[for example, see][]{Thompson95, Lyutikov03}.

We have detected the presence of an emission line-like feature in \src. This detection makes \src\ one of the few magnetars that have shown the presence of emission lines. In the case of SGR 1900+14, an emission feature at 6.4 keV accompanied by a faint hint of its harmonic at $\sim$13 keV was detected during the  first 0.3 s of the precursor of a strong burst \citep{Strohmayer00}. In XTE J1810--197, a narrow 12.6 keV emission feature was reported by \citet{Woods05} by using \textit{RXTE} data during the bright X-ray tail of a burst. In 1E 1048.1--5937, \citet{Gavriil02} reported the presence of a $\sim$14 keV emission feature during the initial stages of a burst detected with \textit{RXTE}. For this magnetar, \citet{An14} also reported a $\sim$13 keV emission feature using \textit{NuSTAR} data, thereby ruling out any instrumental effects. In all these magnetars, the emission line is a transient feature observed occasionally during the burst, but we have found the emission feature during persistent emission in \src.  We did not find evidence for this emission line in the individual spectra of the two brightest  bursts nor in the cumulative burst spectra.

There are several possibilities that could explain the presence of this emission line but, owing to the lack of sufficient data, these possibilities come with many limitations. It is possible that the emission line is a result of the fluorescence of iron due to the presence of relatively cool material near the neutron star. We have detected an emission line with an equivalent width of about $\sim 0.24$ keV. This is similar to the emission line generally observed in accreting X-ray pulsars \citep[e.g.,][]{Naik11, Naik12, Sharma23}. Another possibility for the presence of the emission line is related to proton and alpha-particle cyclotron transitions in astrophysical systems with ultrastrong magnetic fields \citep{Strohmayer00, Ibrahim02, Ibrahim07}. For magnetars having $B\lesssim 10^{14}$ G, the electron cyclotron absorption line energy is $\mathcal{O}$(MeV), which is out of bounds for \astro\ detectors. The proton and alpha-particle cyclotron resonances are well within reach, with fundamentals at $E_p = 6.3 (1+z)^{-1} (B/10^{15} G)$ keV and $E_{\alpha} = 3.2 (1+z)^{-1} (B/10^{15} G)$ keV, respectively. Using $(1+z)=1.31$ for canonical neutron star of $M=1.4 M_{\sun}$ and $R=10$ km, the surface field strength in \src\ comes out to be $\sim 1.3 \times 10^{15}$ G for proton cyclotron resonances which is slightly higher than the dipole magnetic field strength derived from the spin-down measurements by \citet{Younes22} and \citet{Coti21}.

In spite of these possibilities and owing to the fact that the emission line has not been detected with other X-ray missions \citep{Coti21, Younes22}, it is possible that its presence in the \astro\ observation is an instrumental systematic effect. Thus, more data and intensive analysis are required before commenting on its viability.

\section*{Acknowledgements}

This work has made use of data from the \astro\ mission of the Indian Space Research Organisation (ISRO), archived at the Indian Space Science Data Centre (ISSDC). We  thank the LAXPC Payload Operation Center (POC) and SXT POC at TIFR, Mumbai for verifying and releasing the data via the ISSDC data archive and providing the necessary software tools. We have also made use of the software provided by the High Energy Astrophysics Science Archive Research Center (HEASARC), which is a service of the Astrophysics Science Division at NASA/GSFC. This work has also made use of data supplied by the UK Swift Science Data Centre at the University of Leicester. This paper makes use of the following software packages: \textsc{numpy} \citep{Harris20}, \textsc{scipy} \citep{Virtanen20}, \textsc{astropy} \citep{Astropy18}, and \textsc{matplotlib} \citep{Hunter07}. We thank the anonymous referee for insightful comments and suggestions, which helped to improve the manuscript significantly.


\section*{Data Availability}

Data used in this work can be accessed through the Indian Space Science Data Center (ISSDC) at 
\url{https://astrobrowse.issdc.gov.in/astro\_archive/archive/Home.jsp}. 



\bibliographystyle{mnras}
\bibliography{refs} 







\bsp	
\label{lastpage}
\end{document}